# TOWARDS DIRECT NUMERICAL SIMULATION OF TURBULENT CO-CURRENT TAYLOR BUBBLE FLOW

**E.M.A. Frederix[1], J.A. Hopman[1], T. Karageorgiou[1,2] and E.M.J. Komen[1]**

[1]Nuclear Research and Consultancy Group (NRG), P.O. Box 25, 1755 ZG Petten, the Netherlands
[2]Faculty of Technical Mathematics and Informatics, Delft University of Technology, P.O. Box 5, 2600 AA Delft, the Netherlands

**Extended Abstract**

The modeling and simulation of two-phase flows using CFD can contribute significantly to the topic of nuclear reactor safety, e.g., in the accurate prediction of the pressurized thermal shock phenomenon under two-phase conditions. However, there are still many challenges remaining in the development of a general two-phase model which is capable of capturing all two-phase flow regimes within the same setting, ranging from bubbly flow to large interface flow such as slug flow, Taylor bubble flow or stratified flow. One of those challenges is the appropriate representation of the behavior of turbulence at a large scale two-phase interface, and the subsequent break-up of such an interface into smaller structures. This requires the development and validation of more advanced two-phase turbulence models. The first step in this direction was made in [1], in which Egorov's idea of turbulence damping near a two-phase interface was generalized towards a more mesh-independent formulation. For specific settings, the model could be calibrated to give good agreement with experimentally measured velocity and velocity fluctuation profiles of two-phase stratified flow. However, for real-world problems, a much more general interfacial turbulence damping model should be developed which adapts to local conditions automatically.

To achieve this goal, much more insight into turbulent flow near a large two-phase interface should be developed. Direct Numerical Simulation (DNS) of relevant large interface two-phase turbulence has the potential to contribute to this, as it can produce more refined insight while being complementary to experimental data. Therefore, with this goal in mind, in the current study, we present a simulation strategy for DNS of turbulent co-current Taylor bubble flow. This is a continuation of the work presented in [2] in which Large Eddy Simulation (LES) of co-current turbulent Taylor bubble flow was presented. It was observed that one of the main challenges is the physically accurate prediction of the behavior of the Taylor bubble skirt, and the related bubble shedding. An underestimation of the turbulent fluctuations in the wake of the Taylor bubble was observed in the LES results. It was suggested that this is related to over-prediction of the loss of void of the Taylor bubble due to bubble shedding induced by an LES mesh resolution which is not sufficient to capture the break-up and bubble formation accurately. To counter this, in the current work we present a DNS approach of co-current turbulent Taylor bubble flow called RK-Basilisk, based on the Basilisk code with local adaptive grid refinement. This strategy allows for very high mesh resolution near the bubble's interface while elsewhere the grid is allowed to be coarser. Each time step, the mesh is adapted based on a void fraction criterion. Basilisk's underlying data structure which is based on an 'octree' allows for much faster solution procedures and, therefore, a much greater number of grid points as compared to the LES simulations which were performed using the more general OpenFOAM code. We compare the results against experimental data of the same setting, as well as the co-current LES OpenFOAM results.

The setting of turbulent Taylor bubble flow in co-current conditions allows for both lower order turbulence model development and the validation of more general two-phase modeling strategies. Apart from that, Taylor bubble flow in itself also bears relevance to specific two-phase flow situations in nuclear installations, e.g., in emergency core cooling systems during a loss of coolant accident, or in the U-tubes of a stream generator during a pipe rupture. The current work contributes to an advancement in simulation capability for such situations.

## 1 INTRODUCTION

Multiphase flow problems are met not only in several natural phenomena (e.g. sediment transport in river flow, blood flow) but also in many real-world engineering problems in automotive, oil and gas, and power generation sectors. In Nuclear Reactor Safety (NRS), different multiphase flow regimes may arise depending on the volumetric flow rates, the geometry (pipe diameter and length, wall roughness) and orientation of the pipe, and several fluid properties (density, viscosity, surface tension). The three main steps for the modeling of a multiphase flow problem consist of the identification of the type of the multiphase flow, the specification of the physical process together with the phenomena that may occur, and, finally, the determination of the mathematical model. For this reason, several reduced models have been used (e.g., RANS, RELAP5, TRACE, etc.) which, however, are applied to limited situations and cannot cover every case sufficiently accurately [3]. For example, in cases where the interface follows an irregular, unsteady form, high fidelity simulations are needed to capture its behavior accurately.

Among the resulting two-phase flow regimes in NRS, there is the case of slug flow where under certain conditions of fluid flow within a pipe, small bubbles are merged and form large bubbles, known as Taylor bubbles [4]. Taylor bubbles are of bullet shape, occupy most of the pipe diameter and appear in sequences. In NRS, Taylor bubble flow is met in emergency core cooling, or the pressurized thermal shock (PTS) applications.

Many attempts have been made to simulate Taylor bubble flows but the majority of them are limited to laminar flow such as in [5], [6], [7], [8]. According to [5] and [2], despite the number of studies performed, there is a lack of realistic data in the case of turbulent co-current and counter-current Taylor bubble flow. In particular, a fully 3D turbulent co-current Taylor bubble flow needs high-resolution calculations such as DNS or LES for capturing the complex formation of the interface, increasing the demand on computational resources. Such high-resolution calculations are usually not feasible for industrial applications and, therefore, there is a lack of accurate data. The study of [2] tried to fill this gap by performing LES of co-current turbulent Taylor bubble flow using OpenFOAM. Since there were no other available simulation results in the literature, the predictions were validated against the experiments of [9]. However, a deviation of the turbulent fluctuations in the wake of the Taylor bubble due to over-prediction of the loss of void of the Taylor bubble made the authors of [2] conclude that the LES mesh resolution is not sufficient to capture the break-up and bubble formation accurately. Meanwhile, other researchers have also performed high fidelity simulation of turbulent Taylor bubble flow, e.g., see [10] and [11].

Therefore, the main target of the current study is to continue the work of [2] by illustrating a DNS approach of turbulent co-current Taylor bubble flow using RK-Basilisk which is based on the Basilisk code. Basilisk is an open-source software for the solution of partial differential equations describing fluid flow on adaptive Cartesian meshes. Due to Basilisk's "octree" grid approach and its accurate VOF advection scheme, it will be shown that it is a better choice than OpenFOAM for DNS in two-phase flows (via settings of [12] and [9]) since it reduces the computational cost significantly. All in all, the setting of turbulent Taylor bubble flow in co-current conditions allows for both lower-order turbulence model development and the validation of more general two-phase modeling strategies.

The layout of this paper is as follows. In Section 2, the governing equations, the numerical modeling and the methods used for the simulation of the resolved interface in two-phase flows are introduced. Next, in Sections 3 and 4, RK-Basilisk is validated and tested for single phase turbulent channel and pipe flow, and for a laminar rising bubble case against [12]. In Section 5, high-fidelity simulation of turbulent co-current Taylor bubble flow is presented and compared qualitatively and quantitatively with studies of [2] and [9]. Finally, in Section 6, conclusions are presented.

## 2 MODEL AND METHOD

### 2.1 Governing set of equations

To develop a simulation strategy for fully resolved two-phase flow, a common starting point is the Volume of Fluid (VOF) method. The VOF method consists of a momentum equation, continuity equation and scalar advection equation. The form of the incompressible momentum equation which is adopted in this work is given by:

$$\frac{\partial \mathbf{u}}{\partial t} + \nabla \cdot (\mathbf{uu}) = \frac{1}{\rho}[-\nabla p + \nabla \cdot (2\mu \mathbf{D})] + \sigma \kappa \nabla f + \mathbf{g} + \mathbf{h} \tag{1}$$

with velocity vector $\mathbf{u}$, time $t$, mass density $\rho$, pressure $p$, viscosity $\mu$, deformation tensor $\mathbf{D}$, surface tension $\sigma$, interfacial curvature $\kappa$, void fraction $f$, gravitational acceleration $\mathbf{g}$ and $\mathbf{h}$ a forcing term used to model impenetrable solid walls. The deformation tensor is given by

$$\mathbf{D} = \frac{1}{2}[\nabla \mathbf{u} + (\nabla \mathbf{u})^\mathsf{T}] \tag{2}$$

The forcing term $\mathbf{h}$ is computed using a direct volume penalization method [13], and can be expressed as

$$\mathbf{h} = -\frac{H(\mathbf{x})\mathbf{u}}{\tau} \tag{3}$$

with $\tau$ the time scale of velocity damping (taken to be much smaller than the numerical time step $\Delta t$, which will be introduced shortly) and $H(\mathbf{x})$ a masking function which is unity for cells inside the solid wall and zero outside. The forcing term $\mathbf{h}$ is designed to damp any perturbations in $\mathbf{u}$ inside the wall.

Next, the continuity equation is given by

$$\nabla \cdot \mathbf{u} = 0 \tag{4}$$

which implies that the velocity field is divergence free, i.e., the fluid is incompressible. The third equation to be introduced in the VOF method is the scalar advection equation for the void fraction $f$, i.e.,

$$\frac{\partial f}{\partial t} + \mathbf{u} \cdot \nabla f = 0 \tag{5}$$

and describes the evolution of the void fraction $f$ which is, by definition, unity for the first (liquid) phase and zero for the second (gas) phase, while smoothly varying between those two limits at the two-phase interface. Both the mass mixture density $\rho$ and mixture viscosity $\mu$ are algebraically related to their phase-specific counterparts using the following void fraction-based mixing rules:

$$\mu = \frac{1}{\frac{f}{\mu_\ell} + \frac{1-f}{\mu_g}} \tag{6}$$

and

$$\rho = f\rho_\ell + (1-f)\rho_g \tag{7}$$

with $\mu_\ell$, $\mu_g$, $\rho_\ell$ and $\rho_g$ the phase-specific viscosities and densities for the liquid ($\ell$) and gas (g). The harmonic mixing of the viscosity in Eq. (6) is chosen for more numerical stability.

### 2.2  Temporal and spatial discretization schemes

In order to solve the system of equations (1), (4) and (5), we use the open-source code Basilisk [14]. Key to Basilisk is that it is built around an 'octree' concept, which allows for 2-fold refinement of cells in each spatial direction. The three-dimensional base grid consists of eight 'level 2' cells, i.e., a block grid with two cells in each spatial direction. Each of those eight cells can then be refined by another eight 'level 3' cells. In turn, each 'level 3' cell can be refined further to eight 'level 4' cells, and so on. The hierarchy of cells is tracked in a so-called 'octree', providing an efficient method for the construction of computational stencils. The local refinement of cells, which can also occur adaptively in time, allows capturing certain features in the flow with high precision, while leaving other regions relatively unrefined. Such an approach works particularly well for multi-scale problems such as two-phase flow. Two-phase interfaces may be captured very sharply at higher refinement levels while for single phase flow features, typically exhibiting larger length scales, it is sufficient to be discretized on lower refinement levels. Potentially, Basilisk herewith offers a computationally cost-effective platform for the development of a DNS approach for turbulent Taylor bubble flow.

While Basilisk offers full capability of solving the set of equations (1), (4) and (5), we have essentially developed our own Navier-Stokes solver which uses a different strategy in spatial and temporal discretization of the momentum equation. In essence, the work of [15] was followed. Key components of the newly developed solver are the following:

- Whereas the standard momentum equation solver of Basilisk uses a two-point approach with upwinding (see [16]), we opt for the more general PISO-like approach built around a general Runge-Kutta (RK) Butcher tableau [15]. This allows for the use of different RK schemes, depending on the problem at hand. For implicit RK stages, we solve two PISO iterations (i.e., two pressure Poisson equations are solved using Basilisk's standard multigrid solver). For explicit RK stages, only one PISO iteration is solved
- The convective term is discretized using the finite volume approach (i.e., using Gauss' theorem) with a flexible limiter approach. The limiter function $\varphi(r)$ with $r$ the ratio of successive gradients, allows the interpolation scheme used in the convective term to gradually change from upwind ($\varphi = 0$) to linear ($\varphi = 1$). Similar to what was done in [2], a flux limiter (in this case the Minmod limiter [17]) is only applied at computational faces which have a non-zero gradient of $f$ on at least one adjacent cell. Away from the interface (i.e., in single-phase regions) the limiter function is always set to unity, in order to reduce artificial numerical dissipation
- Discretization of the viscous term and pressure gradient is done using linear schemes
- The standard Basilisk discretization and solution algorithm is retained for the treatment of the surface tension force and scalar advection equation using a conservative, non-diffusive geometric VOF scheme [18], [19]
- Special attention is paid to the proper implementation of Basilisk's so-called 'cell and face metrics', allowing for the use of deformed or stretched meshes in single phase flow. Because we rely on Basilisk's standard tools for the solution of the scalar advection equation, stretched meshes could only be used in single-phase problems. Implementing proper cell and face metrics in the two-phase part of Basilisk remains yet to be done
- When using the 'octree' approach, Basilisk's mesh is essentially always a cube which can be stretched in three dimensions into a rectangular prism. While Basilisk offers tools for the modeling of internal boundaries, e.g., for a cylindrical wall necessary for Taylor bubble flow, it is not capable of doing so in conjunction with parallelized code. Therefore, we implemented the direct volume penalization method directly into the Navier-Stokes solver instead

In the remainder of this paper, we refer to our solver as *RK-Basilisk*. Before embarking on more complicated problems, the newly developed solver with the direct volume penalization method was first successfully tested and validated for lid-driven cavity flow. However, presentation of those results is omitted here in favor of a more detailed discussion on more advanced simulation results, to be presented in the next Sections.

## 3 Simulation of single phase turbulent channel and pipe flow

### 3.1 Turbulent channel flow

For proper DNS of two-phase flow, it is first essential to study the performance of our approach in single phase turbulence. To do so, first a turbulent channel flow at $Re_\tau = u_\tau h/\upsilon = 180$ is simulated on a (stretched) computational domain of $(2\pi, 2, \pi)$, similar to the domain used by Kawamura et al. [20]. Here, $h$ is the channel half height, $\upsilon$ is the kinematic viscosity and $u_\tau$ is defined as:

$$u_\tau = \sqrt{\frac{\tau_w}{\rho}} = \sqrt{\upsilon \left(\frac{\partial \langle u \rangle}{\partial y}\right)_{\text{wall}}} \tag{8}$$

With $\tau_w$ the wall shear stress and $\langle u \rangle$ the mean streamwise velocity. As a reference, we use the data of Kawamura et al. [20] which was obtained for both the relatively coarse mesh A (1M cells) and relatively fine mesh E (500M cells). The flow is driven by a pressure gradient in the streamwise direction. The domain has periodic boundaries in the streamwise and spanwise directions. The Crank-Nicholson scheme is used for the temporal discretization.

A Courant-Friedrichs-Lewy (CFL) condition of 0.4 is used to adaptively set the time step size. Several computational grids are used to test performance in relation to mesh resolution. All grids are refined equally in all three dimensions up to a level $l$, where the refinement in each direction is given by $2^l$ cells, here levels 5, 6 and 7 were used. A stretching is applied in the wall normal direction from cells at the wall towards the cells in the bulk. In Table 1, the grid specifics for the three levels of refinement are given, including the achieved stretching factor $g$. Cell dimensions are expressed in wall units, i.e.,

$$\Delta x^+ = \Delta x \frac{u_\tau}{v} \qquad (9)$$

etc.

| Mesh | $l$ | $\Delta x^+$ | $\Delta y^+$ wall | $\Delta y^+$ bulk | $\Delta z^+$ | $N_{cells}$ | $g$ | CFL |
|---|---|---|---|---|---|---|---|---|
| S5 | 5 | 35.34 | 3.00 | 27.19 | 17.67 | 32768 | 1.158 | 0.4 |
| S6 | 6 | 17.67 | 1.50 | 13.85 | 8.84 | 262144 | 1.074 | 0.4 |
| S7 | 7 | 8.84 | 0.75 | 6.99 | 4.42 | 2097152 | 1.036 | 0.4 |

*Table 1: grid specifics of the three used turbulent channel flow grids*

For the level 7 mesh, several temporal schemes were tested, for its influence on solution accuracy. Next to Crank-Nicolson, these included a second order Diagonally Implicit RK (DIRK2) scheme [21], the third order Runge Kutta scheme (RK3) and Midpoint (MP). In Table 2, simulation run times are given for each grid and each temporal scheme, along with the number of processors and number of cells per processor.

| Mesh | Scheme | CFL | $N_{proc}$ | $N_{cells}/N_{proc}$ | Clock time (h) |
|---|---|---|---|---|---|
| S5 | Crank-Nicolson | 0.4 | 8 | 4096 | 0.7 |
| S6 | Crank-Nicolson | 0.4 | 32 | 8192 | 3.7 |
| S7 | Crank-Nicolson | 0.4 | 64 | 32768 | 43.6 |
| S7 | DIRK2 | 0.4 | 64 | 32768 | 76.0 |
| S7 | RK3 | 0.4 | 64 | 32768 | 52.8 |
| S7 | Midpoint | 0.4 | 64 | 32768 | 61.5 |

*Table 2: properties of the performed turbulent channel flow simulations*

The results are assessed in terms of the streamwise velocity profile and Root Mean Square (RMS) velocities. The simulations are run from $t = 0$ through $t = 100$ (in wall units). Results are taken by averaging the fields between $t = 20$ and $t = 100$ and averaging in the streamwise and spanwise direction, resulting in plots as a function of the wall-normal coordinate $y^+$.

From Figure 1, it can be seen that the level 5 mesh performs significantly worse in terms of the flow characteristics. The largest difference between the level 6 and 7 mesh can be seen in terms of the velocity profile, where the level 7 grid gives excellent results very close to Kawamura's reference data. However, even for S7, a significant underestimation of the streamwise RMS is observed, in particular in the region beyond $y^+ = 10$. Figure 2, on the other hand, shows the same profiles of the mean streamwise velocity and the RMSs computed on mesh S7 for four different temporal schemes (Crank-Nicolson, Midpoint, DIRK2 and RK3). While all schemes agree well with the streamwise velocity profile of Kawamura, a significant improvement of the prediction of the streamwise RMS is observed for the DIRK2 and RK3 schemes, underlining that the choice of temporal integration scheme plays an important role in the accurate simulation of turbulent flows.

## 3.2 Turbulent pipe flow

Next, turbulent pipe flow is simulated to validate RK-Basilisk's performance in a cylindrical domain, for which it is necessary to use the volume penalization method in order to model the wall. The domain has streamwise length $2\pi$ and a height and width of $\pi$, such that cells at any given level have an aspect ratio of 2:1:1. The pipe is modeled using the volume penalization method to have a radius of $R = 1$. The domain is coarsened for cells which lay at radii $r > R$, in order to reduce the number of unused cells. Results of Kasagi et al. [22] are used as a reference. The computational domain of Kasagi is similar except the pipe length is 5 diameters. Since RK-Basilisk uses a Cartesian structured grid approach, there is no use for stretching in a cylindrical domain. Therefore, only uniform grids, or grids with local '2 by 2 by 2' refinement at the wall are of interest. Uniform grids of levels 6 through 8 are tested and grids with maximum refinement of level 8 and minimum levels of 5, 6 and 7 are tested. As a post-processing step, the results are interpolated using the nearest neighbor interpolation, to a sufficiently accurate cylindrical mesh. This cylindrical mesh has stretching in the radial direction. Averages are taken over the time interval $t = 20$ through $t = 100$, and in the azimuthal and streamwise directions, leaving the results as a function of the radial direction only. For all simulations, the RK3 scheme is used as it is shown to be the most accurate scheme for turbulent channel flow. Table 3 shows the specifics for the different meshes and simulation runs. Figure 3 shows cross-sectional snapshots of the magnitude of the velocity field of a quarter of the domain, including the wall (in blue), for the 6 different meshes.

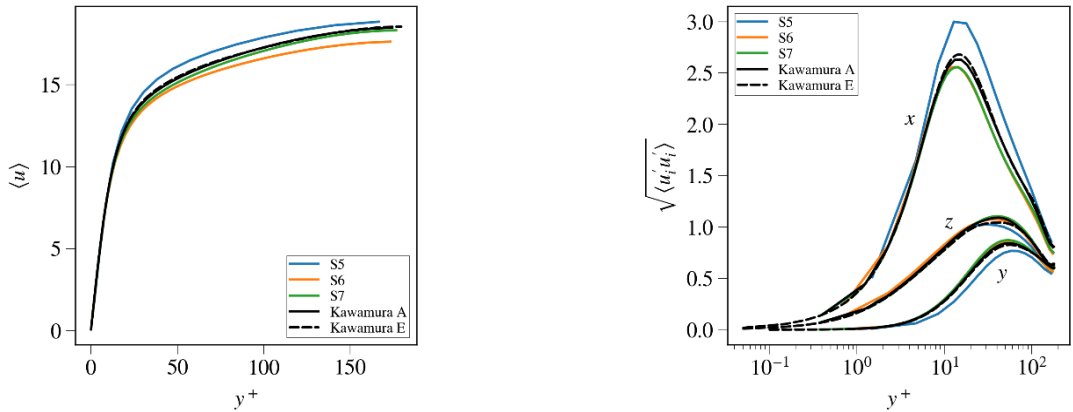

*Figure 1: mean streamwise velocity and RMS velocities of turbulent channel flow computed on three stretched meshes (S5, S6 and S7) using the Crank-Nicolson temporal scheme*

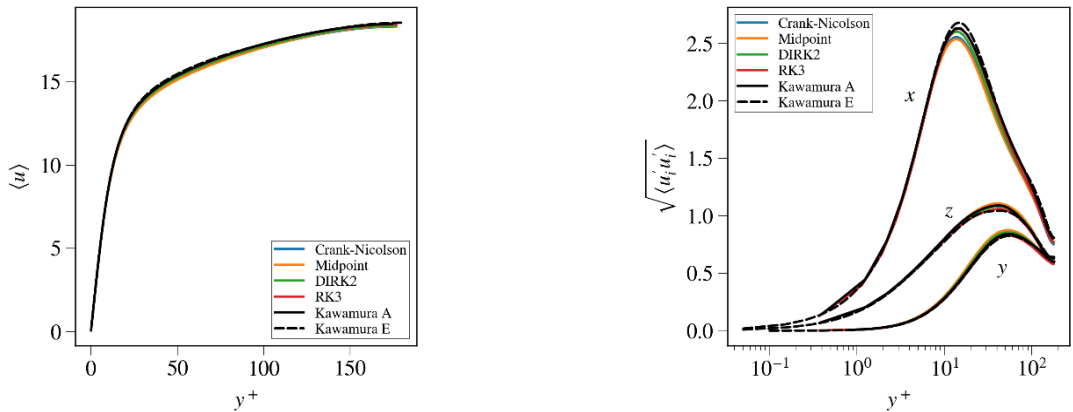

*Figure 2: mean streamwise velocity and RMS velocities of turbulent channel flow computed using four temporal schemes (Crank-Nicolson, Midpoint, DIRK2 and RK3) on the stretched mesh S7*

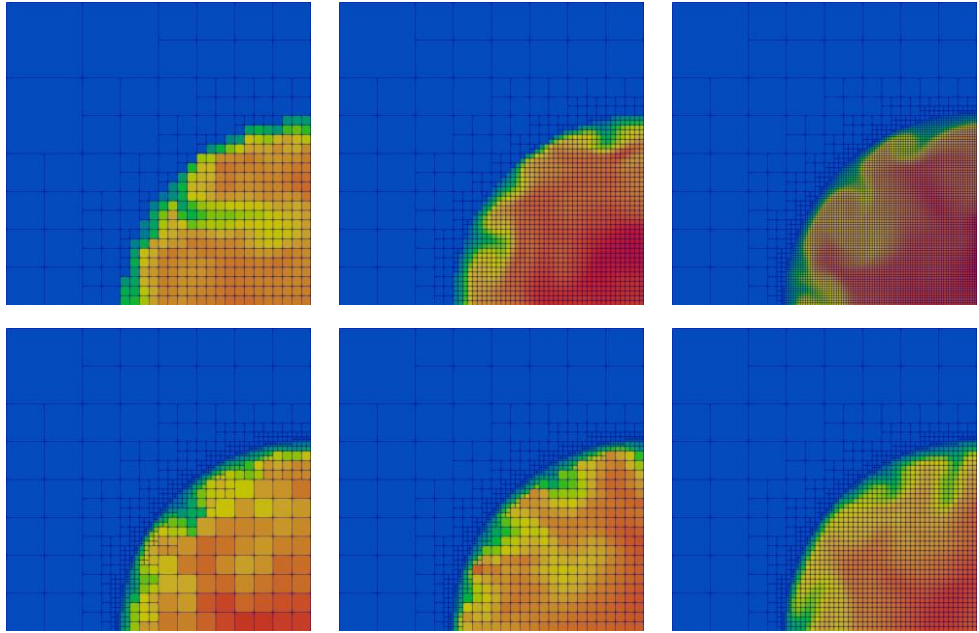

*Figure 3: cross-sectional snapshots of the magnitude of the velocity field of a quarter of the domain, including the wall (in blue), for Mesh U6 (top left), U7 (top center), U8 (top right), R58 (bottom left) R68 (bottom center) and R78 (bottom right)*

| Mesh | $l_{min}$ | $l_{max}$ | $\Delta x^+$ wall | $\Delta x^+$ bulk | $\Delta y^+, \Delta z^+$ wall | $\Delta y^+, \Delta z^+$ bulk | $N_{cells}$ | Scheme | $N_{procs}$ | $N_{cells}/N_{proc}$ | Clock time (h) |
|---|---|---|---|---|---|---|---|---|---|---|---|
| U6 | 6 | 6 | 17.67 | 17.67 | 8.84 | 8.84 | 91k | RK3 | 8 | 11k | 5.8 |
| U7 | 7 | 7 | 8.84 | 8.84 | 4.42 | 4.42 | 687k | RK3 | 32 | 21k | 24.6 |
| U8 | 8 | 8 | 4.42 | 4.42 | 2.21 | 2.21 | 5.4M | RK3 | 64 | 84k | 261.0 |
| R58 | 5 | 8 | 35.34 | 4.42 | 17.67 | 2.21 | 429k | RK3 | 32 | 13k | 42.0 |
| R68 | 6 | 8 | 17.67 | 4.42 | 8.84 | 2.21 | 477k | RK3 | 32 | 15k | 44.4 |
| R78 | 7 | 8 | 8.84 | 4.42 | 4.42 | 2.21 | 980k | RK3 | 64 | 15k | 48.6 |

*Table 3: properties of the performed turbulent pipe flow simulations*

Figure 4 shows the mean streamwise velocity and the RMS velocities of the turbulent pipe flow simulations computed on the three uniform meshes as listed in Table 3. Agreement with the reference data is good, although the mean velocity profile is slightly overpredicted. This may be a result of the lack of conformity of the mesh with the cylinder wall. While mesh U8 produces slightly better results, mesh U7 is remarkably close to the reference data too, with approximately 5% of the computational cost of U8. Figure 5 shows the same data for the refinement meshes. It is shown that mesh R78 produces excellent results in comparison with the reference data. However, mesh R68 deviates quite dramatically, in particular for the RMSs, from the reference data. This suggests that local '2 by 2 by 2' refinement near the wall may have a negative effect on the quality of the prediction in terms of turbulent statistics, and should therefore be applied with care. Nevertheless, mesh R78 compares rather favorably with the result of mesh U8 at only 19% of the computational cost of U8.

In summary, the presented simulation results for turbulent channel and pipe flow show that the developed approach produces excellent results at sufficiently refined computational meshes while the computational cost remains reasonable.

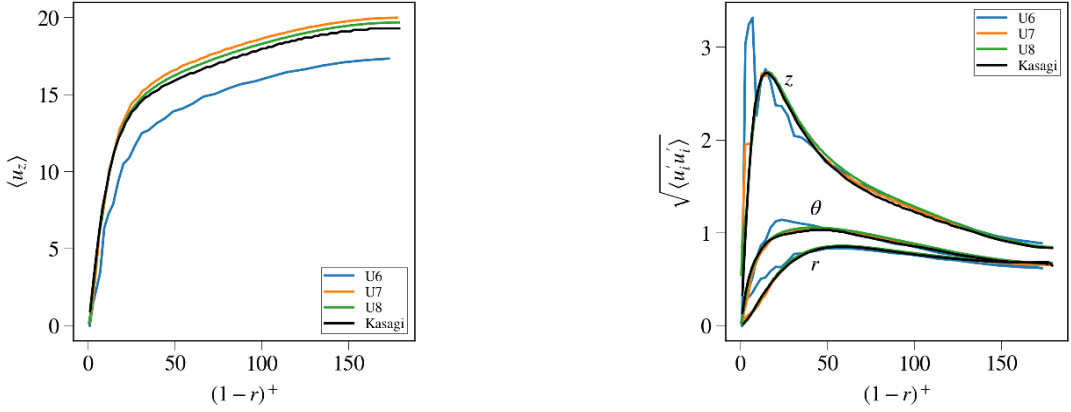

*Figure 4: mean streamwise velocity and RMS velocities of turbulent pipe flow computed on three uniform meshes (U6, U7 and U8) using the RK3 temporal scheme*

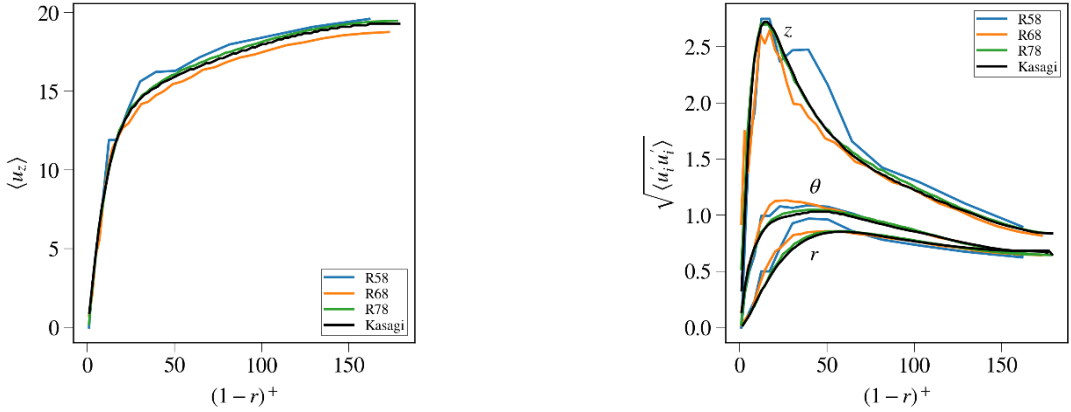

*Figure 5: mean streamwise velocity and RMS velocities of turbulent pipe flow computed on three refined meshes (R58, R68 and R78) using the RK3 temporal scheme*

## 4 Simulation of a laminar rising bubble

In the previous Section, single phase turbulence was studied and simulated. Next, in this Section, we extend the validation study of RK-Basilisk towards two-phase flow, within the two-dimensional rising bubble setting presented by Hysing et al. [12]. We compare the performance of RK-Basilisk against that of OpenFOAM, where we use the OpenFOAM VOF solver as reported in Frederix et al. [2], which also used a RK/PISO approach similar to RK-Basilisk. Simulations are implemented for a two-dimensional bubble which is released in a rectangular box and rises under the influence of buoyancy while undergoing shape deformation. The validation is performed for two different configurations as presented in [12]. In the first case, small density and viscosity ratios with strong surface tension are applied, while in the second case, simulations are performed with high density and viscosity ratios in combination with low interfacial forces. Table 4 indicates all the fluid parameters for the two cases as presented in [12]. Except for gas density and dynamic viscosity and surface tension (and therefore, also the Eötvös number), all parameters are identical in both cases.

| Case | $\rho_l$ [kg/m$^3$] | $\rho_g$ [kg/m$^3$] | $\mu_l$ [kg/m/s] | $\mu_g$ [kg/m/s] | $g$ [m/s$^2$] | $\sigma$ [N/m] | Eo |
|------|---------------------|---------------------|------------------|------------------|---------------|----------------|-----|
| 1    | 1000                | 100                 | 10               | 1                | 0.98          | 24.5           | 10  |
| 2    | 1000                | 1                   | 10               | 0.1              | 0.98          | 1.96           | 125 |

*Table 4: physical parameters of the two rising bubble cases*

The initial setting and the boundary conditions are identical for both configurations and are taken from [12]. In particular, the initial bubble shape is a circle of diameter 0.5 m and the center of the circle is initially placed at the point **x** = (0.5,0.5) of a rectangular domain of size 1 by 2 m. A no-slip boundary condition is set at the top and the bottom and a free-slip boundary condition at vertical walls. In RK-Basilisk, the domain is captured on a 2 by 2 square surface with internal boundaries. The free-slip boundaries are captured by setting a zero viscosity for $x<0$ and $x>1$. For those regions, coarsening of the mesh is applied in order to reduce the number of unused cells. Figure 6 shows the final prediction of the shape of the rising bubble at $t = 3$ s for both cases, computed on mesh E5 which is to be introduced momentarily. Refinement can be observed near the bubble interface while coarsening is shown inside the free-slip walls present in the domain.

Results in [12] are compared both quantitatively and qualitatively. The necessity for quantitative comparison for the validation of the simulation comes from the fact that different codes with identical problem formulations do not always result in the same shapes. In the current flow problem, this occurs mostly in the second case which is the most challenging one. Thus, the visualization of the shape of the bubble may act as an indicator but cannot be considered as the only criterion for drawing conclusions. For this reason, three different benchmark quantities are defined in the study of Hysing et al., which describe both direct and indirect topological parameters: vertical bubble position, bubble circularity and bubble rise velocity. For the purposes of the current study, only the vertical position ($y$) and the rise velocity ($v$) are analyzed for both cases.

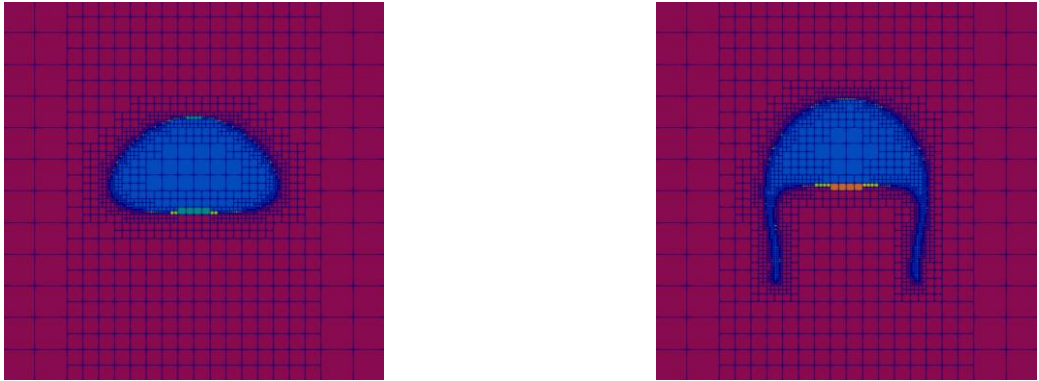

*Figure 6: final shape of the rising bubble at t = 3 s for case 1 (left) and case 2 (right), computed by RK-Basilisk on mesh E5*

| | | | | Case 1 | | | | | | Case 2 | | | | | |
|---|---|---|---|---|---|---|---|---|---|---|---|---|---|---|---|
| | | | | RK-Basilisk | | | OpenFOAM | | | RK-Basilisk | | | OpenFOAM | | |
| Mesh | $N_y$ | $N_{cells}$ | $N_{proc}$ | $E(y)$ | $E(v)$ | $T_{sim}$ [s] | $E(y)$ | $E(v)$ | $T_{sim}$ [s] | $E(y)$ | $E(v)$ | $T_{sim}$ [s] | $E(y)$ | $E(v)$ | $T_{sim}$ [s] |
| A | 32 | 512 | 1 | 0.55 | 3.3 | 6 | 1.0 | 5.5 | 5 | 0.64 | 2.7 | 7 | 2.0 | 9.4 | 5 |
| B | 64 | 2048 | 1 | 0.25 | 0.61 | 31 | 0.65 | 4.1 | 36 | 0.78 | 3.1 | 35 | 1.3 | 6.0 | 40 |
| C | 128 | 8192 | 1 | 0.15 | 0.31 | 315 | 0.51 | 3.6 | 402 | 0.53 | 3.3 | 350 | 0.59 | 3.5 | 439 |
| D | 256 | 33k | 2 | 0.067 | 0.14 | 2.3k | 0.41 | 3.0 | 4044 | 0.32 | 2.2 | 2.8k | 0.29 | 2.0 | 4.2k |
| E | 512 | 131k | 7 | 0.019 | 0.049 | 4.4k | 0.26 | 2.0 | 11k | 0.16 | 1.1 | 3.9k | 0.16 | 1.0 | 14k |
| F | 1024 | 524k | 16 | - | - | 38k | - | - | 60k | - | - | 23k | - | - | 78k |
| E5 | >32 | ~2k | 2 | 0.44 | 2.5 | 603 | - | - | - | 2.3 | 2.0 | 696 | - | - | - |
| E6 | >64 | ~3k | 2 | 0.19 | 1.3 | 1.4k | - | - | - | 3.2 | 1.3 | 1.3k | - | - | - |
| E7 | >128 | ~9k | 2 | 0.098 | 0.53 | 1.9k | - | - | - | 3.1 | 1.9 | 2.4k | - | - | - |
| E8 | >512 | ~34k | 2 | 0.052 | 0.17 | 7.9k | - | - | - | 2.1 | 1.8 | 4.3k | - | - | - |

*Table 5: Simulation data for each mesh, code and case, with vertical position accuracy measure E(y), rise velocity accuracy measure E(v) and simulation clock time $T_{sim}$ in s. The accuracy measures are expressed as a percentage*

For each case, six simulations with uniform meshes have been performed with each code. In addition, four more simulations with adaptive mesh refinement are performed, in which, in each time step, the grid is refined when the change in *f* is above the threshold value of 0.001, and coarsened otherwise. The numerical predictions of the simulations with the most refined mesh of each software are used as reference solutions and therefore, the accuracy of each simulation is tested in the sense of self-convergence. Moreover, both codes are compared to each other and also against the predictions of Hysing et al. [12] (using the MooNMD code). We express the self-convergence in terms of accuracy measures $E(y)$ and $E(v)$. These are defined as the scaled L1-norm of the difference between $y$ and $v$ with the reference $y$ and $v$, respectively, and taken over the time interval [0,3]. The accuracy measures are expressed as a percentage.

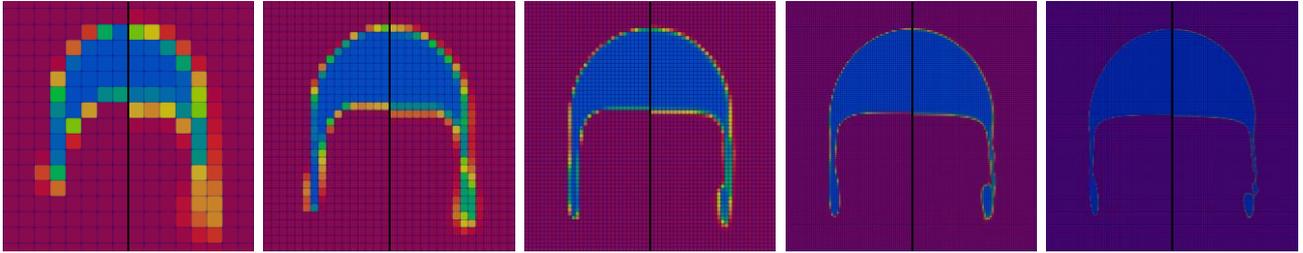

Figure 7: comparison of RK-Basilisk's (left halfs) and OpenFOAM's (right halfs) prediction for the bubble shape at $t = 3$ s for case 2 and, from left to right, mesh A, B, C, D and E

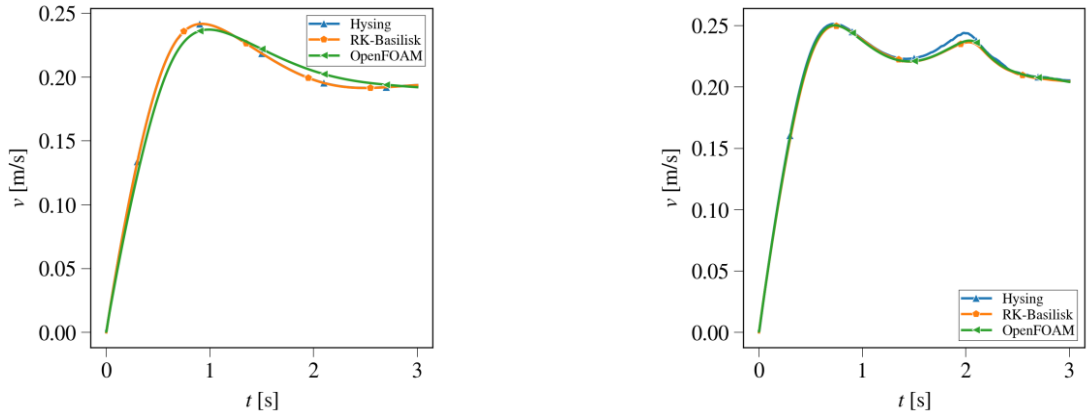

Figure 8: comparison of the bubble rise velocity for Hysing (MooNMD code), RK-Basilisk and OpenFOAM computed on mesh F for case 1 (left) and case 2 (right)

All simulations in both RK-Basilisk and OpenFOAM use a cell-centered grid and the solvers use the same configuration settings and schemes. In particular, DIRK2 is used for the temporal discretization while central spatial discretization is used for the convective and diffusive terms. Moreover, the time step size is adaptively set using a CFL condition of 0.1, while 12 corrector steps are used in the PISO algorithm. This rather small CFL number and large number of PISO correctors was particularly necessary for case 2, in order to reach convergence, due to the relatively large viscosities in conjunction with the only partly (diagonally) implicit treatment of the viscous term in the PISO approach.

Table 5 provides an overview of the used meshes, number of cells in the *y*-direction, total number of cells, number of processors and resulting accuracy measures and computational clock times, for each case and code. The adaptive meshes are designated as mesh E5-E8, where E indicates that the upper level of refinement is equal to the resolution of uniform mesh E, and the number indicates the lower level of refinement (i.e., $2^5$, $2^6$, $2^7$ and $2^8$, respectively). As an illustration, Figure 6 shows the final bubble shape at $t = 3$ s for case 1 and case 2 computed on mesh E5. Next, Figure 7 shows, for case 2, a side-by-side comparison of the prediction of the final bubble shape at $t = 3$ s for RK-Basilisk (left half) and OpenFOAM (right half), for mesh A-E. Finally, Figure 8 shows a comparison of the bubble rise velocity $v$ as a function of time $t$ for RK-Basilisk, OpenFOAM and Hysing, for both cases. From these tables and figures, the following can be observed:

- RK-Basilisk shows better self-convergence to the reference solution (mesh F) on uniform coarser meshes than OpenFOAM as the deviation in the vertical bubble position and bubble rise velocity is always smaller in RK-Basilisk. In particular, as the uniform mesh is becoming finer (after mesh C), the

- difference in absolute values between the two codes becomes more significant, i.e., values in RK-Basilisk are one order of magnitude smaller
- For uniform coarse meshes, computational times are very similar while for larger meshes, RK-Basilisk is up to 3 times faster than OpenFOAM. For adaptive meshes in RK-Basilisk, the time difference further increases
- The adaptive meshes in RK-Basilisk show reasonable qualitative (i.e., shape) and quantitative (within about 3 percent for both the bubble rise velocity and position) agreement with the uniform reference mesh at a lower computational cost
- Figure 7 shows a qualitative agreement between RK-Basilisk and OpenFOAM. However, minor differences are observed. It appears that RK-Basilisk is capable of retaining the interface to one computational cell, whereas the interface in OpenFOAM is somewhat more smeared despite its interface compression scheme. Moreover, OpenFOAM shows a break-up of the bubble's skirt whereas RK-Basilisk predicts a very thin elongated filament while preventing break-up until the end of the simulation ($t = 3$)
- Figure 8 shows excellent quantitative agreement for the bubble rise velocity between RK-Basilisk and Hysing for case 1, whereas OpenFOAM deviates somewhat, even for the very high mesh resolution of mesh F. In case 2, RK-Basilisk matches with OpenFOAM

In summary, the rising bubble simulations presented in this section illustrate well the differences between OpenFOAM and RK-Basilisk. Moreover, results suggest that RK-Basilisk is more accurate and more efficient than OpenFOAM. Improved numerical efficiency is particularly shown for the adaptive mesh refinement approach available in RK-Basilisk. These results, together with those presented in the previous Section on single-phase turbulence, pave the way for the use of RK-Basilisk for turbulent Taylor bubble simulations, as will be presented in the next Section.

## 5 Co-current turbulent Taylor bubble flow

In this Section, simulation results will be presented of co-current turbulent Taylor bubble simulation. As a point of reference, we use the experimental data of Shemer et al. [9] and the OpenFOAM simulation data of Frederix et al. [2]. First, we briefly discuss the simulation strategy, meshing and averaging approach and, finally, present a comparison of the simulation results with the two references.

### 5.1 Simulation strategy, meshing and averaging

We consider a the turbulent co-current Taylor bubble setting of Shemer et al. [9]. In this setting, water is flowing in upward direction in a vertically placed pipe with inner diameter $D = 1.4$ cm and at diameter-based bulk Reynolds number Re = 8250 which gives a superficial bulk velocity of approximately 0.59 m/s. The diameter-based shear Reynolds number is approximately $Re_\tau = 532$. An air Taylor bubble is released in the flow. Due to buoyancy, the Taylor bubble attains a velocity of approximately 0.83 m/s with respect to a frame of reference fixed to the wall of the pipe. In a frame of reference attached to the bubble, measurements can be done of the velocity in the wake of the Taylor bubble. Generally, a toroidal vortex can be observed in the wake, which pushed liquid towards the Taylor bubble tail in the axial center of the pipe while pulling liquid away from the Taylor bubble tail near the wall.

To simulate this setting, the simulation strategy presented in [2] is followed here. For details, the reader is referred to that reference. However, for completeness, a short summary is given here of the main features of the approach:

- Simulations are performed in a Moving Frame of Reference (MFR) attached to the Taylor bubble in a domain which is 16$D$ long
- Ahead of the Taylor bubble, single phase liquid turbulence is recycled over an interval of $\pi D$ at the top boundary of the pipe. The imposed liquid velocity is scaled in such a way that the bubble's nose remains fixated in the MFR while the wall velocity is adjusted in such a way that the superficial liquid velocity is 0.59 m/s
- A precursor simulation is performed for developed single-phase turbulence at Re = 8250. A snapshot of the velocity field is used as an initial condition for the two-phase simulation

- A Taylor bubble is initialized in such a snapshot, as a cylinder with length $2D$ and a diameter at 85% of the pipe diameter, together with a hemispherical head. The center of the cylinder is initially located at a distance of $(\pi+3)D$ away from the top boundary of the domain. After some time, the bubble can be seen to slightly elongate
- The simulation is ran for 0.3 s (which corresponds to roughly one flow-through time for the bubble in the $16D$ long section of the pipe), after which temporal averaging of the solution fields is started
- Four runs are performed on the same mesh, each with a statistically independent initial velocity field. These runs are then ensemble-averaged to obtain statistically converged results
- As a post-processing step, fields are averaged in the circumferential direction, yielding surfaces of data in the $(r,z)$-plane

In [2] it was observed that the loss of void of the Taylor bubble was relatively large. For the coarse grid used there, a loss of void of about $1D$ in 0.8 s was observed while for the finest grid the loss of void reduced to about $1D$ in 2.4 s. In the current RK-Basilisk simulations, the loss of void is significantly reduced as compared to [2]. Therefore, for all meshes, an averaging time of 2 s could be used while setting $N_{ensemble} = 4$. I this way, statistics are gathered in a window of about 8 s, which corresponds to about 30 bubble flow-through times in a $16D$ long pipe. Table 6 shows the three different meshes that are used for the RK-Basilisk simulations. Also shown is the ensemble-averaged loss of void decay rate, expressed in $D/s$. It can be seen that even on the coarsest mesh, the loss of void remains well within the $D/2$ threshold value which was used in [2] in the averaging time span of $T = 2$ s. Note that the cell sizes of the selected meshes are such that they are far from DNS quality. Near the wall, additional refinement is necessary to capture the boundary layer sufficiently accurately. However, these meshes may already give good insight into the performance of RK-Basilisk at reasonable computational cost.

| Mesh | $l_{min}$ | $l_{max}$ | $N_{cells}$ | $N_{cells}/D$ min | $N_{cells}/D$ max | $\Delta x^+$ min | $\Delta x^+$ max | $N_{ensemble}$ | $T$ [s] | CFL | Decay rate [$D/s$] |
|---|---|---|---|---|---|---|---|---|---|---|---|
| 9-10 | 9 | 10 | ~950k | 32 | 64 | 16.6 | 8.3 | 4 | 2.0 | 0.4 | 0.11 |
| 10-10 | 10 | 10 | 3.5M | 64 | 64 | 8.3 | 8.3 | 4 | 2.0 | 0.4 | 0.057 |
| 10-11 | 10 | 11 | ~5.4M | 64 | 128 | 8.3 | 4.2 | 4 | 2.0 | 0.4 | 0.016 |

*Table 6: co-current turbulent Taylor bubble simulation meshes and data*

For all simulations, a max CFL criterion of 0.4 is used. The convective term is discretized using the two-phase limiter scheme as presented in Section 2.2, in which the Minmod limiter is applied only near the two-phase interface while reverting to a central scheme elsewhere. For the time integration, we use the second-order IMEX scheme presented in [21], which is family of the DIRK2 scheme used in previous Sections of this paper. It was found that an explicit treatment of the convective term reduced the computational time somewhat, while remaining sufficiently accurate and stable.

## 5.2 Results

Figure 9 shows a cross-sectional impression of the full instantaneous solution of co-current turbulent Taylor bubble flow computed on mesh 10-10. Compared to [2], much less formation of smaller bubbles is observed in the wake of the Taylor bubble. A strong toroidal vortex can be seen in the wake of the bubble, where a positive current is directed towards the Taylor bubble in the core of the pipe (red) while close to the wall the direction of the streamwise velocity changes to the left due to liquid, being pushed downward by the rising Taylor bubble (blue). Ahead of the Taylor bubble, the single-phase turbulent flow appears as relatively undisturbed even at small distances ahead of the bubble's nose.

Figure 10 shows a space, time and ensemble-averaged cross-sectional view for the three used meshes. The average shape of the bubble is outlined with the thick black line, using an iso-contour at $\langle f \rangle = 0.5$. The largest loss of void is observed in mesh 9-10, although the difference is rather small with mesh 10-10 and 10-11. Quite interestingly, the average shape of the bubble tail varies from flat (9-10) to somewhat concave (10-10) and somewhat convex (10-11). The streamlines indicate that the bubble itself consists of a large toroidal vortex with its rotational center relatively close to the tail. This was also observed in [2]. However, the prediction of the size of the toroidal vortex in the wake of the bubble is slightly larger than what was observed in [2], in particular for mesh 9-10. Due to the reduced loss of void as compared to [2], streamlines seeded inside the bubble and outside the bubble do no longer significantly overlap.

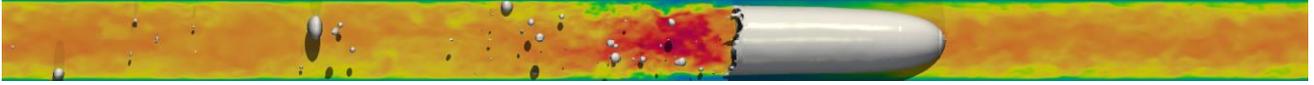

*Figure 9: cross-sectional impression of the full instantaneous solution (cropped for r < D/2) of co-current turbulent Taylor bubble flow computed on mesh 10-10 with RK-Basilisk. Color indicates the streamwise bubble velocity relative to the Taylor bubble with blue being negative and red being positive*

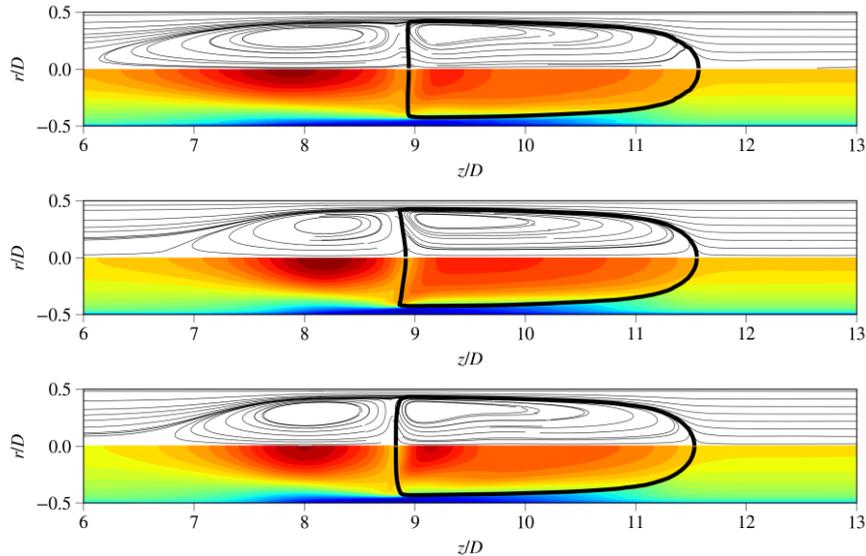

*Figure 10: space, time and ensemble-averaged impression of the Taylor bubble simulation on mesh 9-10 (top), mesh 10-10 (middle) and mesh 10-11 (bottom). The top half of each plot illustrates the streamlines of the mean velocity while the bottom half shows the mean streamwise velocity component in the MFR attached to the bubble*

Figure 11 shows radial plots of the average streamwise and radial velocity in the wake of the Taylor bubble at three distances (0.6$D$, 1$D$ and 2$D$) below the Taylor bubble. Shown are the predictions of the three RK-Basilisk grids, as well as the OpenFOAM simulation data of Frederix et al. [2] and the experimental data of Shemer et al. [9] which, unlike [2], is not corrected to have the right superficial liquid velocity. Generally, the results for mesh 9-10 deviate slightly from those of mesh 10-10 and 10-11. There is qualitative agreement between OpenFOAM and RK-Basilisk, although some minor differences can be observed. Close to the Taylor bubble at 0.6$D$, OpenFOAM predicts a higher streamwise velocity towards the tail of the Taylor bubble. Also, at 1$D$, OpenFOAM shows a stronger radially inward motion of the liquid, confirming the observation that the toroidal wake vortex in the OpenFOAM simulation is shorter than in RK-Basilisk.

## 6    Conclusions

This paper presents a systematic approach for the simulation of turbulent co-current Taylor bubble flow using the RK-Basilisk code platform. A new two-phase solver was developed and extensively tested first for single-phase turbulence and later for two-phase rising bubble flow. For both settings, excellent agreement with data from literature was obtained. DNS quality was achieved for the turbulent statistics in turbulent channel flow and pipe flow. The rising bubble simulations showed that the bubble position, shape and rise velocity were predicted very accurately, at moderate computational cost due to the adaptive refinement strategy. This illustrates well the effectiveness of Basilisk's 'octree' approach. As compared to OpenFOAM, the two-phase interface was shown to be captured always on only one cell, without any artificial smearing. This appeared to reduce break-up of the thin bubble skirt in Hysing's case 2 [12], while maintaining a thin but stable slender filament. These observations paved the way towards the simulation of fully turbulent co-current Taylor bubble flow. On three computational meshes with cell sizes ranging from 16 to 4 wall units, simulations were performed of co-current Taylor bubble flow. Although these meshes are too coarse for DNS quality, they assist in providing further insight into the performance of RK-Basilisk at reasonable computational cost. The same simulation strategy for co-current Taylor bubble flow was used as in [2], allowing for direct comparison of the results. It was observed that the loss of void of the Taylor bubble was much smaller than what was predicted by OpenFOAM. This increased the allowable averaging time while decreasing the number of simulations in the ensemble. We showed that RK-Basilisk predicts a somewhat shorter toroidal vortex in the wake of the Taylor bubble, while results are generally in qualitative agreement with OpenFOAM. However, agreement with the experimental data of Shemer et al. [9] is still relatively poor.

Nevertheless, the current paper demonstrates a 'proof of concept' of simulating turbulent co-current Taylor bubble flow at affordable numerical cost and with good numerical accuracy. With additional computational resources available, this work should pave the way towards DNS of turbulent co-current Taylor bubble flow. Such simulation data may be leveraged for the both validation of lower order models or the development of such models.

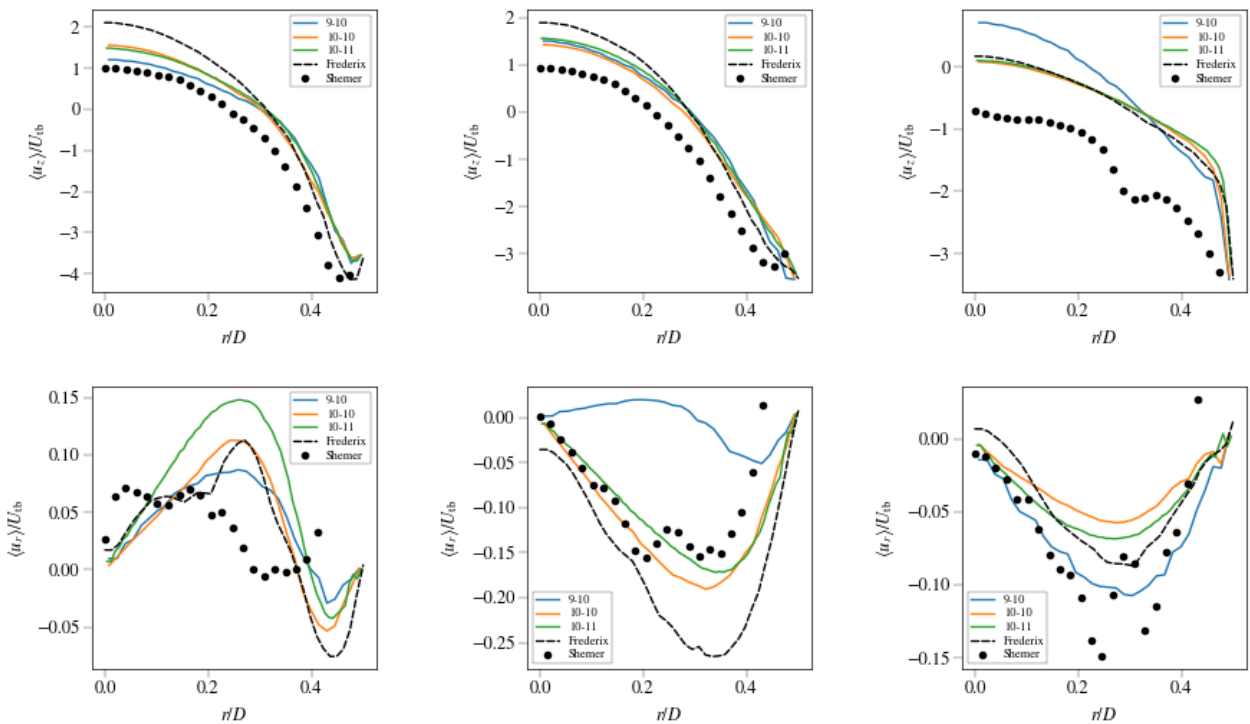

*Figure 11: streamwise (top) and radial (bottom) velocity profiles at 0.6D (left), 1D (center) and 2D (right) below the Taylor bubble tail, computed on mesh 9-10, 10-10 and 10-11. Also shown are the experimental data of Shemer et al. [9] and the numerical data of [2]*

**REFERENCES**

[1] E. Frederix, A. Mathur, D. Dovizio, B. Geurts and E. Komen, "Reynolds-averaged modeling of turbulence damping near a large-scale interface in two-phase flow," *Nuclear Engineering and Design,* vol. 333, pp. 122-130, 2018.

[2] E. Frederix, E. Komen, I. Tiselj and B. Mikuz, "LES of turbulent co-current Taylor bubble flow," *Flow, Turbulence and Combustion,* pp. 1-25, 2020.

[3] E. Frederix, B. Mikuž and E. Komen, "LES of a Taylor bubble in co-current turbulent pipe flow," *Springer, Cham,* no. ERCOFTAC Workshop Direct and Large Eddy Simulation, pp. 135-141, 2019.

[4] D. RM and T. G, "The Mechanics of Large Bubbles Rising Through Liquids and Through Liquids in Tubes," *Proceedings of the Royal Society of London,* vol. Series A. 200, pp. 375-390, 1950.

[5] A. Morgado, J. Miranda, J. Araújo and J. Campos, "Review on vertical gas liquid," *International Journal of Multiphase Flow,* vol. 85, p. 348–368, 2016.

[6] J. Araújo, J. Miranda, A. Pinto and J. Campos, "Wide-ranging survey on the laminar flow of individual Taylor bubbles rising through stagnant newtonian liquids," *International Journal of Multiphase Flow,* vol. 43, pp. 131-148, 2012.

[7] G. Montoya and E. Baglietto, Resolved interface of taylor bubble simulations to support eulerian multiphase closures derivation, 2016.

[8] H. Shaban and S. Tavoularis, "Detached eddy simulations of rising Taylor bubbles," *International,* vol. 107, pp. 289 – 300,, 2018.

[9] L. Shemer, A. Gulitski and D. Barnea, "Experiments on the turbulent structure and the void fraction distribution in the Taylor bubble wake," *Multiphase Science and Technology,* vol. 17, no. 1-2, 2005.

[10] M. Zimmer and I. Bolotnov, "Slug-to-churn vertical two-phase flow regime transition study using an interface tracking approach," *International Journal of Multiphase Flow,* vol. 115, pp. 196-206, 2019.

[11] M. Zimmer and I. Bolotnov, "Exploring Two-Phase Flow Regime Transition Mechanisms Using High-Resolution Virtual Experiments," *Nuclear Science and Engineering,* pp. 1-13, 2020.

[12] S.-R. Hysing, S. Turek, D. Kuzmin, N. Parolini, E. Burman, S. Ganesan and L. Tobiska, "Quantitative benchmark computations of two-dimensional bubble dynamics," *International Journal for Numerical Methods in Fluids,* vol. 60, no. 11, pp. 1259-1288, 2009.

[13] R. Mittal and G. Iaccarino, "Immersed boundary methods," *Annual Review of Fluid Mechanics,* vol. 37, pp. 239-261, 2005.

[14] S. Popinet, 2020. [Online]. Available: http://basilisk.fr.

[15] E. Komen, E. Frederix, T. Coppen, V. D'Alessandro and J. Kuerten, "Analysis of the numerical dissipation rate of different Runge-Kutta and velocity interpolation methods in an unstructured collocated finite volume method in OpenFOAM," *Computer Physics Communications,* p. 107145, 2020.

[16] J. Bell, P. Colella and H. Glaz, "A second-order projection method for the incompressible Navier-Stokes equations," *Journal of Computational Physics,* vol. 85, no. 2, pp. 257-283, 1989.

[17] P. Roe, "Characteristic-based schemes for the Euler equations," *Annual review of fluid mechanics,* vol. 18, no. 1, pp. 337-365, 1986.

[18] G. Weymouth and D. Yue, "Conservative volume-of-fluid method for free-surface simulations on cartesian-grids," *Journal of Computational Physics,* vol. 229, no. 8, p. 2853–2865, 2010.

[19] J. M. Lopez-Herrera, A. M. Ganan-Calvo, S. Popinet and M. A. Herrada, "A vof numerical study on the electrokinetic effects in the breakup of electrified jets," *International Journal of Multiphase Flows,* vol. 71, pp. 14-22, 2015.

[20] H. Kawamura, K. Ohsaka, H. Abe and K. Yamamoto, "DNS of turbulent heat transfer in channel flow with low to medium-high Prandtl number fluid," *International Journal of Heat and Fluid Flow,* vol. 19, no. 5, pp. 482-491, 1998.

[21] U. Ascher, S. Ruuth and R. Spiteri, "Implicit-explicit Runge-Kutta methods for time-dependent partial differential equations," *Applied Numerical Mathematics,* vol. 25, no. 2-3, pp. 151-167, 1997.

[22] N. Kasagi, K. Horiuti, Y. Miyake, T. Miyauchi and Y. Nagano, "Establishment of the Direct Numerical Simulation Data Bases of Turbulent Transport Phenomena," 1993. [Online]. Available: thtlab.jp/DNS/PI12__PG.WL1.

[23] C. E. Brennen, Fundamentals of multiphase flow, Cambridge University Press, 2005.